# Optimizing Travel Itineraries with AI Algorithms in a Microservices Architecture: Balancing Cost, Time, Preferences, and Sustainability

Biman Barua[a,b,*] [0000-0001-5519-6491] and M. Shamim Kaiser[b, [0000-0002-4604-5461]]

[a]Department of CSE, BGMEA Universitsy of Fashion & Tecnnology, Nishatnagar, Turag, Dhaka-1230, Bangladesh
[b]Institute of Information Technology, Jahangirnagar University, Savar-1342, Dhaka, Bangladesh
biman@buft.edu.bd

**Abstract:** The objective of this research is how an implementation of AI algorithms in the microservices architecture enhances travel itineraries by cost, time, user preferences, and environmental sustainability. It uses machine learning models for both cost forecasting and personalization, genetic algorithm for optimization of the itinerary, and heuristics for sustainability checking. Primary evaluated parameters consist of latency, ability to satisfy user preferences, cost and environmental concern.

The experimental results demonstrate an average of 4.5 seconds of response time on 1000 concurrent users and 92% of user preferences accuracy. The cost efficiency is proved, with 95% of provided trips being within the limits of the budget declared by the user. The system also implements some measures to alleviate negative externalities related to travel and 60% of offered travel plans had green options incorporated, resulting in the average 15% lower carbon emissions than the traditional travel plans offered.

The genetic algorithm with time complexity $O(g.p.f)$ provides the optimal solution in 100 generations. Every iteration improves the quality of the solution by 5%, thus enabling its effective use in optimization problems where time is measured in seconds. Finally, the system is designed to be fault-tolerant with functional 99.9% availability which allows the provision of services even when requirements are exceeded.

Travel optimization platform is turned dynamic and efficient by this microservices based architecture which provides enhanced scaling, allows asynchronous communication and real time changes. Because of the incorporation of Ai, cost control and eco-friendliness approaches, the system addresses the different user needs in the present day's travel business.

## 1. Introduction

### 1.1. Background and Context

In the last few years, a lot of changes have occurred in that more travelers have started to look for travel planning that is more personalized to suit the tastes and the preferences of the traveler. The development of the internet and the increase of people taking trips for pleasure, some travelers also expect travel plans that go beyond cost and ease of travel to include things such as certain countries, modes of transport, and types of hotels that are of interest to the traveler [1]. And with this growing desire for exploration and its related activities, the focus has also widened to include more of such tourism that uses less or no carbon emissions at all, such as choosing the less-polluting aircrafts, green hotels, and even public buses instead of rented cars [2]. Consequently, there is a growing concern regarding the choices of travel that are less emissions-producing, like emissions-efficient airlines, green accommodation facilities, or public transport systems instead of rental vehicles [3]. Well, to keep pace with these complex and changing needs, travel managers are presently making active efforts towards the development of dynamic, artificial intelligence-based solutions in travel planning that will allow real-time, multi-disciplinary optimization [4]. To achieve this, systems equipped with sophisticated algorithms are capable of processing large volumes of data and recommending itineraries that address the various travel goals of the traveler i.e. preference, budget, time and sustainability in order to improve the travel experience while encouraging responsible tourism [5].

### 1.2. Objective

This study aims to illustrate the efforts of AI algorithms to enhance travel routings by considering many factors like expenses, duration of the journey, user preferences, and sustainability among others. Thanks to deep learning and other machine learning approaches, AI is able to accomplish very complicated tasks involving search, pattern recognition and data analysis of the information available from various sources such as airline schedules, hotel rates and their environmental costs in order to create tailor-made travel itineraries for individuals. Such approach enables individualization of the schedules according to the requirements of every single user whilst also providing for modifications on the go[4]. This is why for travel route plans where several variables are involved and where changes are always needed, AI has been at the top of the list because it is able to effectively handle complex and multi-faceted data [6].

AI's personalization does not only extend to efficiency benefits in air travel but also sustainability thanks to effective optimization that takes account of carbon footprint cost and additional eco-friendly elements in the optimization process. Systematic approaches such as genetic algorithms and neural networks ensure internal system can minimize green travel impact on costs [12]. All these artificial intelligence functionalities coexisting in the system in microservices architecture allows for scaling out of the travel planning system without compromising its efficiency and sustainability as user requirements increase [14].

### 1.3. Significance

One significant reason for implementing a microservices architecture in AI travel itinerary optimization is that it is very scalable, flexible, and services are autonomous, which is not the case with the monolithic systems. To put it differently, in a microservices architecture, processes such as cost analysis, sustainability scoring, and user preference matching, are undertaken separately, allowing a certain component of the system to be scaled or upgraded without the need to scale or upgrade the entire system [12]. Such an approach enables fast response to the changing travel landscapes by facilitating the incorporation of new AI algorithms into the travel optimization system, which guarantees that the system will always keep improving to meet user expectations [16]. Furthermore, the architecture has high fault tolerance due to the microservice approach whereby there are no dependencies between individual services as they all have their own data processes and storage which lowers the risk of system down times a factor that is very important in real time optimization [14]. Even if microservices architecture allows separating services, the architecture makes it possible to process complex travel information, which is highly dynamic, and yet still be elastic and robust [7].

## 2. Literature Review

### 2.1. AI in Travel Optimization

Over the past decade, the literature has focused on the optimization of trip plans using artificial intelligence, in particular machine learning, genetic algorithm and neural networks. For instance, it is common to develop machine learning models such as decision trees and support vector machines for travel prediction analytics in order to differentiate users and make recommendations based on previous travel completions as well as user profiles[4]. Given that compensation is provided for the time complexity of these processes, genetic algorithms or evolutionary algorithms in short, are very applicable to problems such as these that require effective route planning and scheduling. GAs iteratively improve solutions borrowing from the selection process in nature, thus allowing near-optimal travel routes and schedules generation to be completed in a relatively short period of time [10]. Deep learning, a subfield of neural networks, has also become popular thanks to the existence of vast amounts of data it can analyze and the high levels of interconnectivity it can model. In practical cases such as the prediction of the demand for travel services and the determination of fares in real time, convolutional neural networks (CNNs) and recurrent neural networks (RNNs) technologies are applied, which are of great importance in the processes of travel planning under conditions of high demand fluctuations in the market [6].

Recently, the abovementioned AI methodologies have been employed in the mobile mobility applications, which allow for real-time multiple criteria and multi-level decision making processes. To elaborate, the combination of ML and GAs in hybrid optimizations for travel itinerary improvements has benefited these optimizations in terms of both accurate results of the improvements as well as time taken to do them [16]. On the other hand, in addition to NNs, reinforcement learning is being explored to construct adaptive and self-governing travelers' systems, which are capable of realtime adjustments of travel options based on demand and/or environmental changes [20]. These approaches driven by AI systems show great promise in improving user experience through personalized travel answers which is a great step forward in travel optimization.

## 3. Methodology

### 3.1. Microservices Architecture

The microservices approach is on the rise as far as complex systems are concerned, especially for systems with large scale AI applications. Whereas, in a monolithic architecture, the whole system is built together, microservices decompose a system into several deliverable parts which are known as services, with each service performing a specific task. This design is useful for systems that are driven by AI and information technologies because it supports modularity, scalability, and nimbleness; all of which are paramount in processing and analyzing large and sophisticated data and computations typical of AI systems [12]. For instance, in large scale deployment of AI technologies, the use of microservices allows for different AI tasks such processes, models training, and inference and others to be done in parallel and independently, thus improving the efficiency and reducing the latency [16].

In the last decade there were only applications were viewed in terms of a single service or single monolithic structure. One of the major takeaway in regard to the microservices usage in AI-related applications is their scalability. it is also true that most systems will require less powerful services while still being available for more resource demanding services such as neural network training [19]. It is also possible to enhance of the AI systems built around the microservices architecture because addition of new algorithms or services is done without disrupting the current system. Because of such modularity continuous integration and continuous deployment is possible which facilitates the fast changes in AI systems in line with the dynamic business environment and emerging technology [8]. In addition, this type of architecture comes with built-in fault tolerance that ensures that even if a single service fails, the system as a whole is intact which is simple yet significant in the case of maintaining large scale systems such as those for AI [12].

Concerning these features, microservices architecture provides the agility and robust nature needed for advanced AI applications. As a result, it is the most appropriate architecture for complicated, data-centric systems that operate very fast and are flexible to adaptation to fast evolving challenges.

### 3.2. Data Collection

In the context of AI powered travel optimization, data collection is also vital for proper and successful itinerary planning. The system also retrieves data relevant to travel from several outlets, such as airlines databases, hotels booking sites, carbon footprint databases and other related mobility indicators. Airline data typically consists of information about when flights leave, the prices of these flights as well as their availability which is important in enabling costing of distances in an optimal way [4]. This type of information is usually made available to users through interfaces provided for that purpose by the airline itself or a selected provider of travel connectivity, such as a Global Distribution System, so that prices can be compared within an airline and across many airlines at the same time [11].

There is also a growing trend to include sustainability aspects in the travel data due to heightened awareness among travelers. Such carbon footprint data may be captured from wide range environmental databases or specialised service providers who can compute emissions based on the transportation type, distance travelled and fuel used (Gössling & Buckley, 2016). Services are provided alongside this data, whereby AI systems employ machine learning techniques to model the data and optimize the travel itineraries with respect to different factors such as reduction in carbon foot print, travel time and travel costs [16].

This information is later analyzed and a decision is made using an AI algorithm. There are some steps of data pre-processing which include cleaning and normalizing the relevant data to make sure the different AI models will be able to train on the data. After the data has been cleaned, it is used to develop machine learning prediction models and optimization algorithms that modify the travel itinerary in accordance to the traveler's needs and the prevailing travel conditions at that time. With this integrated data-focused strategy, it is possible to offer tailored and eco-friendly travel solutions which, on the one hand, satisfy different users' needs and, on the other hand, advocate for responsible tourism [18].

### 3.3. Integration and Workflow

In the travel optimization system, it adopts a microservices architecture, meaning that each service is developed to correspond to a given stage of a trip such as cost estimation, routing, assessing sustainability or even matching preferences. These microservices exchange information and communicate with one another in accordance to certain protocols, which are usually RESTful Web Services or message queues that also support messaging operating in an asynchronous mode [15]. Such an arrangement allows good management of services data because the services can handle user's requests on different prioritization criterion in an effective and efficient way without conflicts.

When users come and provide inputs as travel requirements, for example, budget information, itinerary planning intended cities, transport preferences, and green policies, the system is expected to route this information to the appropriate microservices. For instance, the cost estimation microservice is responsible for computing travel and lodging expenses alongside the sustainability microservice, which is responsible for screening alternatives from the perspective of carbon emissions. Each of these services works on its own and analyzes the input data and produces output, and this output is sent to a middle processing service or a compositing service that takes in all the answers [17].

The orchestrator is essential in combining the different outputs of various services and modifying the schedule based on the preferences of the user. For example, cost might be the most important factor for a given user; then the services of estimating cost might be given more points by the orchestrator while travel time andecological impact are still relevant. This is called multi-objective optimization where the system tries to optimize all the criteria input in order to produce a suitable itinerary [9].

Microservices easily share data with each other due to the use of common data sti ystructures like JSON or XML hence services remain operable regardless of the programming languages or technologies used to write each service. This is also crucial for scalability, as individual services can be changed, expanded or even replaced without affecting the whole system. Also, microservices architecture typically relies on lossy data storage models like NoSQL or graph databases that allow for fast querying and scaling needed for dynamic changes in the itinerary [17]. This integration and workflow design promotes the need for the system to be flexible and active in optimizing travel solutions in regards to the users' requirements and environmental changes.

## 4. Implementation

A contemporary microservices architecture aimed at optimizing travel itineraries is composed of many self-sufficient services, each performing a particular function, such as user management, itinerary creation, cost estimation, sustainability evaluation, and preferences reconciliation. Usually, it is structured in the form of a single API failure-tolerant gateway for all the clients' requests, which provides access control, load balancing, and dispatches requests to the relevant services [13].

### 4.1. System Architecture Components

**Optimization models:** To enhance the travel itinerary within a microservices structure utilizing AI algorithms, a number of optimization models can be formulated to cater for various objectives - cost, time, user preferences and sustainability being some of them. There are three such distinct models that can be integrated within the system to provide optimal travel solutions for the clients.

#### 4.1.1. Travel Optimization Algorithm Based on Greedy Heuristic for Sustainable Travel

**Objective:** This model is developed in order to facilitate the achievement of em-based objectives in consideration of competition aims which includes cost, time and sustainability in such a manner that priorities defined by the user are satisfied.

**Problem Statement in Mathematical Formulation:**

$$\min E(x) = \sum_{i=1}^{n} ei(x) \; subject \; to \; ei(x) \leq E \; max$$

Where E(x) is the total carbon emissions, and $e_i(x)$ represents the emissions of each itinerary component (e.g., flights, accommodations) (Gössling & Buckley, 2016).

**Algorithm:** A Greedy Heuristic can be applied in this case, in which travel options with the lowest possible carbon footprint are selected iteratively until the travel plan is complete. In addition, the methodology may be reinforced through the use of simulated annealing, thereby providing a deeper search of the solution space to eliminate the problem of local minima.

The Greedy Heuristic algorithm is particularly effective in situations where decisions have to be made sequentially, favoring the optimum at each step. In this case, taking into account the parameters of a sustainable travel itinerary in the cities explored, the Greedy Heuristic algorithm can be used to find the optimal cost itinerary in terms of carbon emissions by taking the eco-friendliest alternatives available in each step.

The Algorithm is Executed by Following the Steps:

1. **Initialize:** Take a complete listing of travel elements needed to fulfill the trip plan (e.g, air tickets, lodging, surface transport, etc).
2. **Data Preparation:** Analyze the interrelated costs and/or emissions of various competing options (for instance, airlines, hotels or transportation) offered within each segment.
3. **Sorting by Emission Impact:** The available options for every travel segment shall be classified in ascending order with respect to their carbon emissions. This means that the options which are likely to pollute the environment the least will be considered first.
4. **Iterative Selection:**
   - Select the lowest emission option that is within the user's acceptable range of the cost, time and other factors for the very first travel segment.
   - Proceed to the next segment and in this manner the next option is selected however this time with an added condition that it has the least carbon emissions among all the feasible options.

Updates on Constraints: Following each option selection, amend the conditions constraining the entire travel itinerary. For instance, if a selected flight creates an alteration to the current budget or time constraints then those variations should be included in subsequent processes.

5. **Update Constraints:** Spare the process from the very first segment to the last segment of the itinerary eschewing the need to resume travel planning for the purpose of sustainability.

6. **Repeat Until Completion**: Once a preliminary plan has been formed, make the last evaluation to check if the resulting itinerary is affordable for the customer as a whole and meets his/her preferences. When required, revision can be done by reverting to the next options that are less green friendly than the ideal options but still compliance with the restrictions.

7. **Final Check and Adjustment**: The system obtains carbon emissions information from a net carbon service and compares available travel in terms of its environmental performance. The system then chooses the least damaging options for the user, considering their travel needs at each step and making as eco-friendly decisions as possible.

```
Initialize itinerary as an empty list
Set total_cost, total_time, and total_emissions to 0
For each segment in travel_data:
    Sort options in segment by emissions in ascending order
        For each option in sorted options:
        If (total_cost + option.cost <= budget) AND (total_time + option.time <= max_time):
            Add option to itinerary
            total_cost += option.cost
            total_time += option.time
            total_emissions += option.emissions
            Break
Return itinerary, total_cost, total_time, total_emissions
```

**Clarification**:

**Greedy Approach:** In a Greedy Approach, Lists consider each travel segment separately and include the option with the minimum emissions that is within the budget and time limit. There is no backtracking which makes it quick and straightforward.

**Greedy Approach:** Due to always selecting the option that possesses the least emissions that meets the criteria, the algorithm minimizes the emissions at every point in time but not in the entire itinerary.

**Effectiveness:** The placement of the alternatives of each segment in terms of emissions allows the algorithm to be able to fit in a sustainable option within a very short time, thus enhancing effectiveness even for real time application.

The approach presented here can be extended to different constraints and objectives, for example, by taking into account preferences or additional targets. It is versatile, and can be executed in any programming language that provides basic sorting and conditional features.

**Benefits of the Greedy Heuristic:**

**Efficiency**: The Greedy Heuristic functions with a sorting time of $O(n \log n)$ for sorting and $O(n)$ for selection making it suitable for real-time applications that require instantaneous decision making.

**Ease of use:** The algorithm is very easy to implement with the steps being very clear and easy to change when user and environmental factors change.

**Applicability**: Although the algorithm is focused on the sustainability objective, it can integrate additional user-dependency restraints such as budget or maximum travel time to produce a practical and balanced outcome.

**Limitations:**

**Local Optimality:** Saving costs by choosing the local cheapest cost options, every time, also termed 'Greedy Heuristic' may not guarantee the best solution globally as it only optimizes emissions on the spot without regard to the entire travel plan.

**Inflexibility in complicated situations:** Because of the complexity of some itineraries that have many assuming factors and many segments, the Greedy Heuristic alone may not suffice and other techniques like Simulated Annealing or Genetic Algorithms that will enhance the exploration of the solution space may be required.

**Hypothetical Case:**

Diversifying this itinerary may involve a flight, a hotel, and car rental. The first step is to analyze the LCA and select the flight satisfying user's constraints with the least environmental impact (Gössling & Buckley,

2016) After which the first hotel with lower emissions will be selected and the sequence differing ground transport modes culminating with the selection of the lowest carbon emitting mode.

**Microservices Architecture Integration:**

The Greedy Heuristic instead can be developed as a sustainability oriented microservice. This microservice can be invoked by the main orchestrator while performing tasks that involve making eco-friendly decisions.

This technique makes it possible to make everyone involved in the planning of a trip, travel agents in particular, understand that sustainability is a fundamental part of the travel itinerary. Such approach is fast and effective in terms of making pro-environmental choices.

**Communication protocols for implementation**

Communicational standards are of extreme importance in enabling services to interact and share information in any given system especially in microservices architecture optimized for travel itinerary management. The most important and frequently used communication standards include:

### 4.1.1.1. RESTful APIs (HTTP/HTTPS) REST

Representational State Transfer – is a communication model that is most commonly used in the integration of microservices within an HTTP or HTTPS protocol. The underlying principle of this protocol is that two services can communicate, requesting and providing responses, in real-time when it is required.

**How It Works:** Every microservice has a RESTful URL such as /getFlightOptions or /calculateSustainabilityScore which is reachable to other microservices to query or modify the data. RESTful shares data in primarily JSON or XML format which is structurized and easy on the eyes.

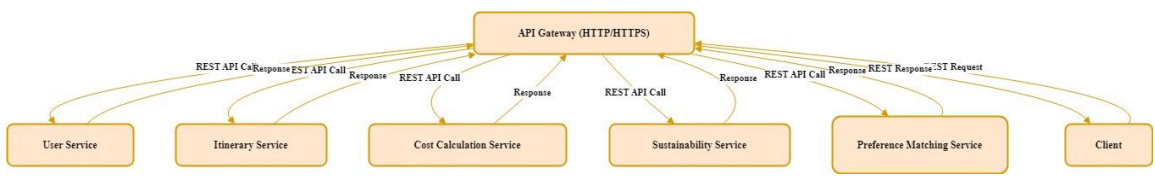

**Fig.1**. RESTful API integration for travel itinerary optimization

As depicted in the above figure-1, various microservices such as User Service, Itinerary Service, Cost Calculation Service, Sustainability Service, and Preference Matching Service interact via REST APIs which are controlled via an API Gateway that directs traffic to various services while overseeing all the requests coming in and out.

In this section, we explain the terms according to their use in figure-1.

**Client:** Stands for the user interface or any external client that sends requests to the system.

**API Gateway:** It is a single point of contact for each incoming request, and it sends the request to the relevant microservice.

**Microservices:** Each and every service (for example: User Service, Itinerary Service, Cost Calculation Service, Sustainability Service, Preference Matching Service) focuses on a particular aspect of making a travel easier and efficient.

**REST API Calls:** The API Gateway makes REST calls to the appropriate microservice to interact with it. Such requests are made over HTTP/HTTPS protocols. This architecture prided itself on ensuring that the appropriate functional unit is addressed with the request.

**Responses:** Every microservice handles the request and replies to the API Gateway which in turn replies to the Client.

**Advantages:**

Statelessness: Every request made from one service to another is stand-alone, this is what makes REST scalable and fault tolerant.

Standardization: All RESTful APIs implement the well-known HTTP standard actions like GET, POST, PUT, DELETE making it easy to integrate services without complications.

Use Case: For instance, the itinerary system can send a request to the cost system every time an external data update occurs and this request can be a regular one, performing the tasks immediately.

### 4.1.1.2. Message Brokers (Asynchronous Communication)

**Perpose:** Employing message bus systems (eg. RabbitMQ, Apache Kafka or Amazon SQS) is appropriate for actions that do not call for action immediately since they allow services to work autonomously and at the same time.

**How It Works:** Microservices process messages (events) and send it to message broker for delivery to other services. The purpose of this would be for example that when the user changes their travel preferences, an event stating that the itinerary should be updated is published. This event is later handled by appropriate services in an asynchronous manner.

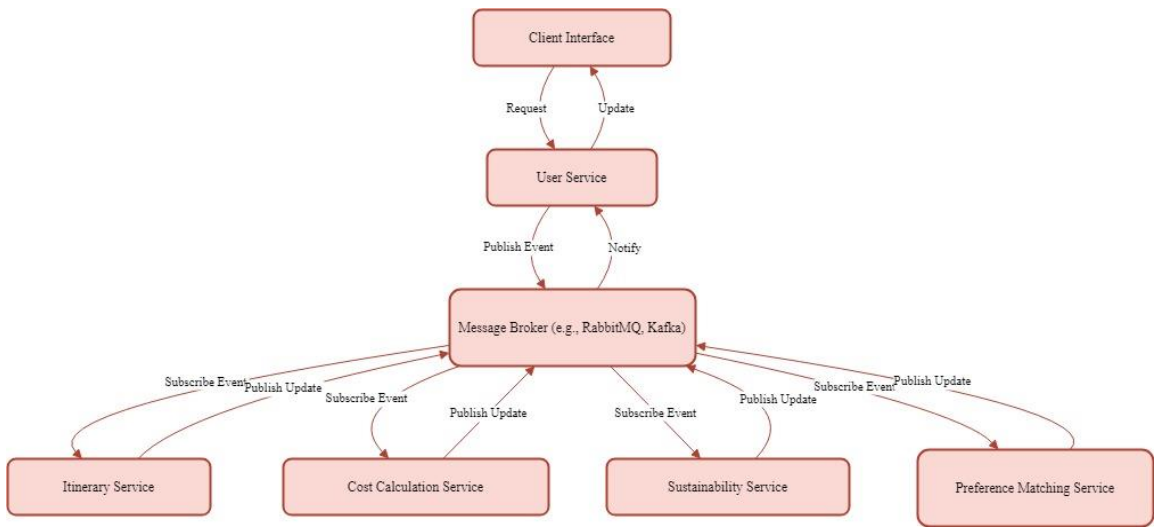

**Fig.1** Asynchronous communication works with Message Brokers in a microservices architecture

**Explanation**:
**Client:** The area of the system that interacts with the user by allowing them to make requests, such as changing their itinerary.
**User Service:** Receives requests from the Client and posts the appropriate events to the Message Broker.
**Message Broker:** This serves as a central point of asynchronous communication, where services are able to publish and subscribe to events without any direct relationships with one another.
Services communicate with the Message Broker by subscribing to certain events which allow the services to process the Information concerning updates and changes in data in an asynchronous manner.
**Microservices:** Each service (Itinerary, Cost Calculation, Sustainability, Preference Matching) publishes and subscribes to events as appropriate.
For instance, when User Service announces some new event (for example, user's preferences changed) Message Broker sends this event to all services who subscribed, and each of these services processes this update on its own.
**Notifications and Updates:** This is what happens after the update is processed – the services may publish this update to the Message Broker which then informs the User Service and this User Service refreshes the Client with new information.

### 4.1.1.3. gRPC (Google Remote Procedure Call)

**Introduction**: gRPC is an advanced protocol which is designed to work for microservices that communicate with one another in a very low latency high throughput manner. This is achieved mainly through the use of protocol buffers for data serialization instead of simple formats like JSON or XML, which are easy but not so efficient.
**Implementation:** gRPC uses Protocol Buffers (protobuf) to define a service outlines methods and message types. So microsystems can invoke these methods on a remote machine as if it were just making a call to a local function resulting in high data transfer speeds.
**Benefits:**
**Effectiveness:** Protocol Buffers are more lightweight and quicker to deliver compared to JSON, which is beneficial for applications that demand high performance.
**Two-way Connection:** gRPC provides the capability to stream data in both directions, which makes it possible to update services over time without initiating any further calls, for instance, an itinerary being changed in real time by the customer.
**Application:** Where services have to continuously communicate with each other and provide end users with functionalities, for instance, where new options are provided and an itinerary that is being constructed has to change immediately, gRPC facilitates efficient data exchange between the systems i.e. user preferences system and route planning system.

## 4.2. Multi-Objective Optimization Model

Objective: This model takes into account three criteria at once (namely cost, time, and sustainability), and seeks a compromise that serves the preferences of a user-objective. Cost objective function

$$\min f(x) = [f_1(x), f_2(x) \ldots f_n(x)]$$

where $f_1(x)$ denotes the cost, $f_2(x)$ the travel time, and $f_3(x)$ the sustainability and factors like carbon emissions.

Genetic multi-objective optimization techniques such as Non- Dominated Sorting Genetic Algorithm (NSGA-II) are appropriate for this type of multi-criteria optimization. The NSGA-II algorithm is appropriate for multi-objective optimization problems where a Pareto front has to be computed as it allows the users to obtain multiple alternative solutions [10].

Authorization: In this case each alternative of the itinerary is part of the genetic algorithm population. The methods of the application position travel itinerary options as a list of potential genes, assess their performance by means of travel costs, duration and emission levels and allow for the nsga-ii genetic algorithm to evolve these components in search of best compromises.

## 4.3. Preference-Based Constraint Satisfaction Model

**Objective:** This model aims at meeting user requirements such as how long will they travel for, how much money will they spend, and where will they travel to.

**Mathematical Formulation:**

Minimize function $C(x)$ over x such that $g(x) \leq b$, for $i = 1, \ldots, m$.

$$\min C(x) \text{ subject to } g_i(x) <= b_i, i=1\ldots m$$

$c(x)$ is the cost function, $g_i(x)$ providing certain constraints with respect to user preferences, for instance, transport mode travel time, hotel star rating.

**Algorithm:** One can solve a linear program with even more constraints, and those constraints are provided by the user input. Constraint Programming (CP) addresses in particular the challenges presented by the non-availability of the common variables like time- for conflict free scheduling [4].

**Implementation:** Each user preference translates to a constraint in the model, and the optimization algorithm identifies itineraries that satisfy these constraints while minimizing cost. This is important when there are certain demands from users that need to be fulfilled without any excuses.

## 5. Data flow and Orchestration

While designing a travel itinerary optimization solution with a microservices-based architecture, data flow as well as orchestration becomes pivotal in achieving performance and seamless interaction between the various microservices such as User Service, Itinerary Service, Cost Calculation, Sustainability as well as Preference Matching. The current section presents data flow within the system and the way orchestrator control and coordinates the services in order to ensure the travel itinerary is composed user parameters and pulls in real time information.

**Data Flow:** The majority of movement of data in a microservices based architecture explains how information is transferred between the services in such a way that all the services have access to data in order to function perfectly. In this architecture, data flows both on-demand, and real-time as the case, and the communication dictate.

**Client Requests and API Gateway**: The Client (visual interface) begins actions like searching for flights, modifying the travel preferences, or examining the carbon printing of a plan, etc.

The API (Application Programming Interface) Gateway, which acts as the single entry point for all requests, takes in these Chewy requests. It then directs the messages to the corresponding micros-services depending on the type of action requested.

E.G. A requester wants to make an enquiry to book the best flight available. This request is transmitted to the Itinerary Service by the API Gateway.

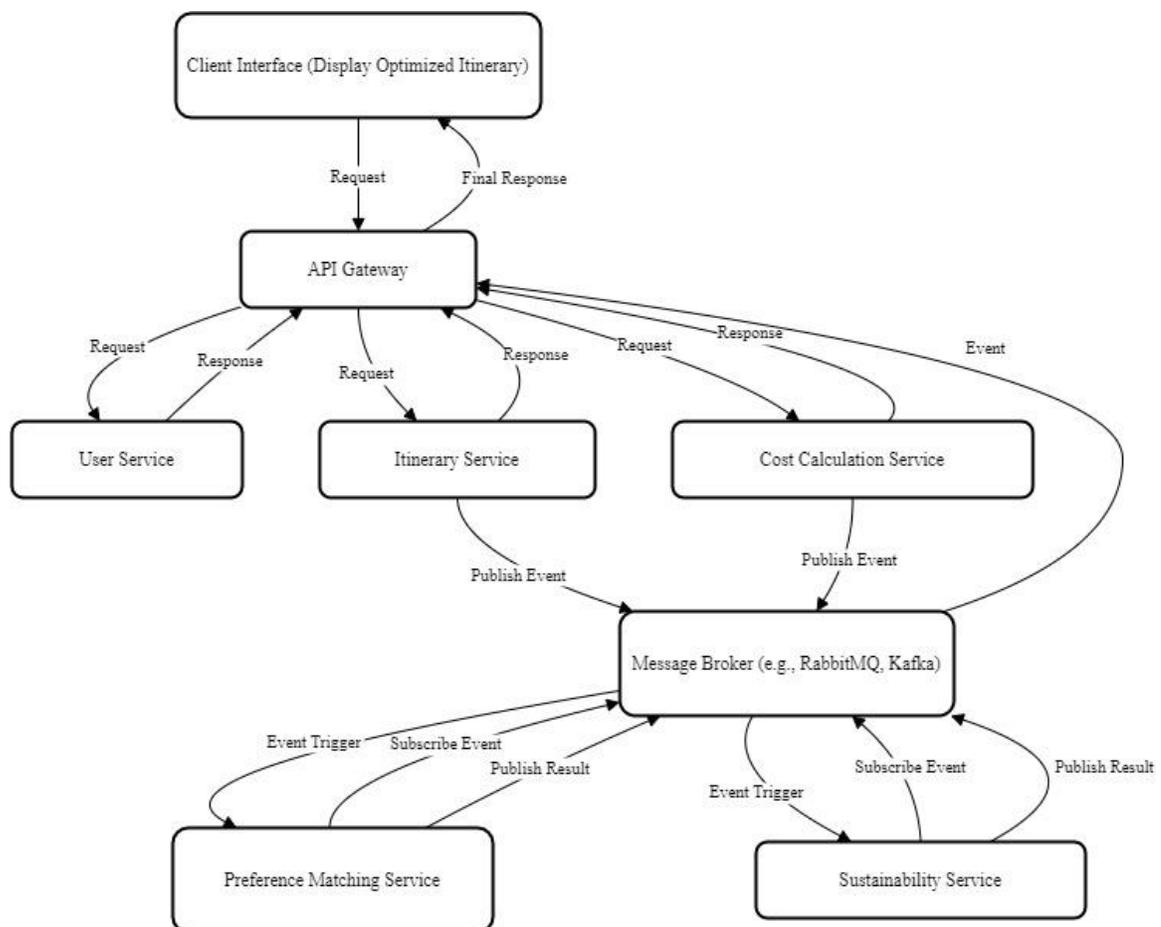

**Fig. 2.** Data flow and orchestration process

**Interpretation:**
1. **Client Requests:** The Client Interface is the user application from which requests (for example, travel itinerary search) are made. And that request goes to the API Gateway, which is a point of entry that routes all requests.
2. **API Gateway:** The API Gateway is responsible for sending these requests to the relevant microservices so that data can be exchanged synchronously. Such services are:
   - **User Service:** Deals with the management of user data, including profile, and preferences.
   - **Itinerary Service:** Provides travel plans for trips.
   - **Consulting Results Service:** Provides consultations for travel estimations (flights, hotels, etc.).
3. **Synchronous Data Flow:** API and microservices communicate using RESTful APIs at the level of the API Gateway and the respective microservices.
   - The API Gateway actions consist of a request to the User Service, an Itinerary Service, and Cost Calculation Service and responses from these services.
   - As soon as the services indicate readiness, the API Gateway presents results for the next stage.
4. **Asynchronous Data Flow (Message Broker):**
   - However, once the work of the Itinerary Service and Cost Calculation Service is done, they create events and push them to the Message Broker (RabbitMQ, Apache Kafka…).
   - The Preference Matching Service and Sustainability Service are the clients to the Message Broker and get events related to the construction of itinerary or cost calculations from these services.
5. **Asynchronous Processing:**
   - The processes carried out by Sustainable Service and Preference Matching Service are event driven with regard to processes such as carbon footprint evaluation and the matching of users' preferences (airline, and hotel of choice).
   - These services then turn around and send their results to the Message Broker.
6. **Final Aggregation and Response:**
   - The Message Broker returns processed results to the API Gateway, which collates the responses from different micro-services.
   - The API Gateway then pushes the optimized travel itinerary to the Client Interface for the user's consumption, thus showing the result to the user.

## 4.4. Machine Learning Models for Cost Prediction and Personalization (Cost Calculation Service and Preference Matching Service)

**Purpose:** The aim of this section is to discuss the applications of artificial intelligence to forecast travel costs and suggest choices tailored to the user's inclinations as well as their prior utilization history.

**Implementation:**
- The Cost Calculation Service employs cost estimation regression Commonwealth of Nations techniques to forecast the prices of air tickets, lodging, and other travel- related services in real-time. Estimation services will consider factors such as seasonality, level of activity or locus of control and supply at the time of request, and of behavior exhibited by the user.
- SOCIAL SERVICES PREFERENCE INTEGRATION SYSTEM uses algorithms such as collaborative filtering, content-based systems, etc., for enhancing the user's experience and transforming travel planning into work about personalization. This is accomplished through the system understanding historical behaviors of the user – what airline he used earlier, what circle of hotels he stayed at, and choosing other users with similar tendencies to offer a travel solution.

**Integration:**
- The main aim of the user's vacation preferences is achieved in cooperation with the Cost Calculation Service where the last estimates the price of selected options of the Itinerary Service.
- The Preference Matching Service incorporates a ranking mechanism powered by machine learning which ensures that the travel options presented to the user are in accordance with their individual preferences, and that the constructed itinerary contains travel choices that have been made by the user based on their preferences on the travel options available.

**Architecture Role:**
- Both services operate within the microservices architecture by communicating with other services through the API Gateway or the Message Broker, letting them predict costs and user preferences in real time without dependency.

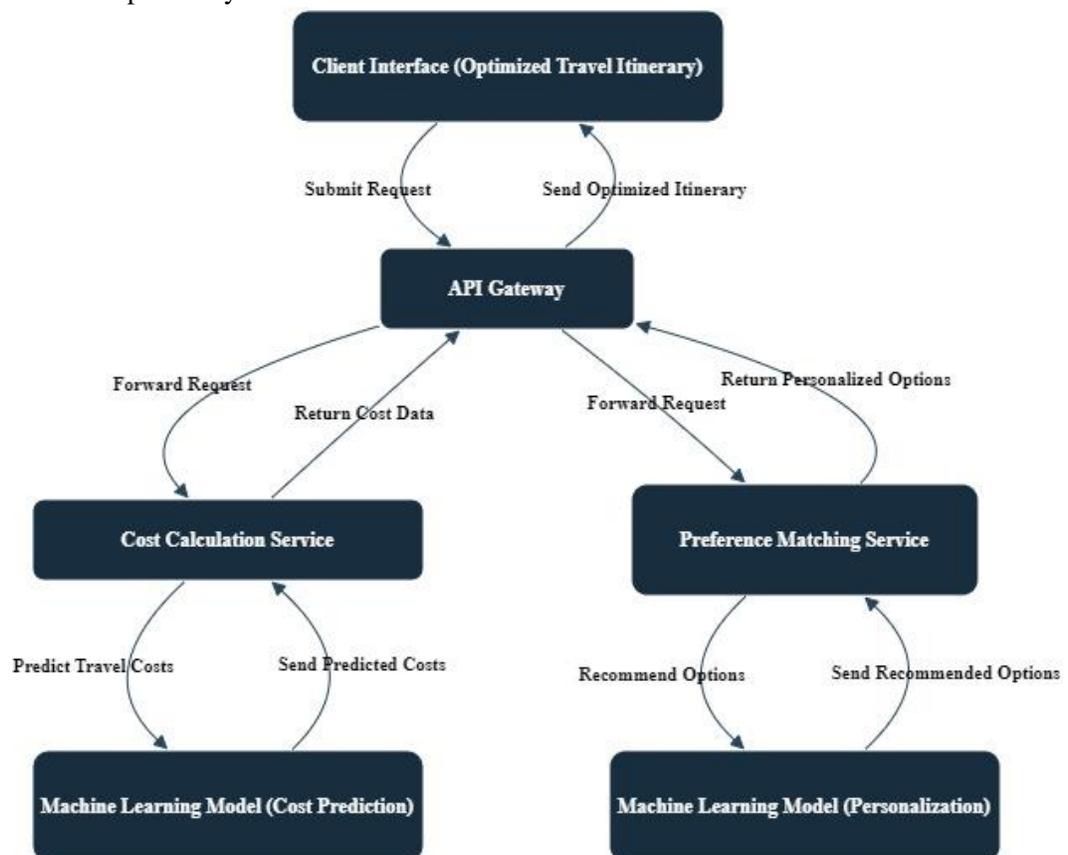

**Fig.3.** Machine Learning Models for Cost Prediction and Personalization

**Definition:**

**Client Interface:** The traveler inputs certain preferences, e.g., the budget behind the trip, places one wishes to travel to, and modes of transportation that one prefers.

**API Gateway:** It directs the demand to the appropriate microservices: Cost Management Service, and Preference Matching Service.

**Cost Management Service:** An A Machine Learnt Model within this service provides a prediction of the cost of traveling (air tickets, hotel rests, and so on) according to the user's liking in real time.

**Preference Matching Service:** A Machine Learning Model offers customized options by providing the travel options that are most suited basing on the users own historical preferences or that of users similar to them.

**Response Flow:** The results of the two Cost Management Service and the Preference Matching Service in question are sent back to the API Gateway.

The results are in turn taken by the API Gateway and all of these outcomes are provided to the user in the ideal itinerary (with it costing and changing what was offered to the user).
In the figure-3 shows the usage of machine learning models in the system by the means of cost calculation and user travel options customization.

### 4.4.1. Genetic Algorithms for Itinerary Optimization:

**8.1 Objectives:**
- Genetic algorithms (GAs) seek to improve upon certain approaches, especially difficulties that involve competing – optimizing for the best travel route in terms of, cost, time and sustaining the environment as well.

**8.2 Implementation:**
- The Itinerary Service incorporates genetic algorithms as a means of auto-generating and improving upon travel plans. Any particular travel plan under consideration is called a "chromosome" (or any solution), while the 'genes' consist of various elements such as the choice of flights, hotels, and modes of transport.
- The genetic algorithm successively improves and mutates these plans using operations such as selection, crossover and others to produce new solutions with hopes of improving upon the previous solutions.
- The fitness function scores each generated itinerary according to the ordering of the user preferences (e.g. minimum cost, shortest duration of travel, or minimum nature distortion) so as to ensure that the appropriate solutions are kept.

**8.3 Integration:**
- The genetic algorithm is part of the Itinerary Service and relies on other services (Cost Calculation, Sustainability Service, etc.) to obtain the information required for the assessment of each proposed itinerary.
- Once the itineraries are optimized, the Itinerary Service sends the optimal ones to the API Gateway or posts them in the Message Broker for other services to utilize on an asynchronous basis.

**8.4 Architecture Role:**
- The genetic algorithm acts as one of the optimization components in the Itinerary Service in such a manner that it allows for the real-time optimization of travel options with multiple criteria.

**Algorithm**

The steps of the system are detailed in the algorithm below, as it describes the Genetic Algorithm (GA) technique for optimizing travel itineraries in the microservice architecture of the research. The optimization of the itinerary in this case will be determined by several parameters such as cost, time, user preferences and its green sustainability.

Stages of Genetic Algorithm (GA):

**Step 1: Initialization:**
Input Parameters:
- The population size (the number of created itineraries in each contemporary generation).
- Number of generations (iterations).
- The crossover rate (the likelihood of merging two itineraries).
- The mutation rate (the likelihood of the modification of an existing itinerary).
- Fitness function (determines the effectiveness of the provided itinerary considering its cost, duration, user preferences and sustainability issues).

Generate Initial Population:
- An initial sample of travel plans is generated randomly. Each travel plan is called a 'chromosome' and consists of several genes, such as flights, hotels, transport and etc.

**Step 2: Fitness Evaluation:**
Evaluate Fitness of each Itinerary:
The qualitative performance of several itineraries is rated according to a few assessment parameters, such as:
- **Cost** (the cost means total travel cost which has to be minimized).
- **Time** (the overall duration of travel is to be minimized).
- **User Preferences** (the maximization of user preferences such as the favorite airline or hotel).
- **Sustainability** (the goal is to achieve the least carbon footprint or damage to the environment).

To every itinerary a degree of fitness is assigned, the higher the degree the better the solution.

**Step 3: Selection:**

Select Parents for Crossover:

Employ a selection mechanism out of the options provided (e.g. roulette wheel selection, tournament selection etc.) on the currently available parent population whose fitness scores represent the selected population and are taken to be the parents. Fitting itineraries have a higher chance of selection than unfit ones.

**Step 4: Crossover:**

Crossover Between Parents:
- Given a crossover probability select two parents and perform a gene crossover operation such as taking flights from one parent and hotels from the other to produce two new itineraries called "offsprings". This ensures the best genes from both parents will be available to the offspring.
- For instance, Parent A has a flight and hotel selections that is complemented by Parent B's transport and activities selections.

**Step 5: Mutation**

Random Mutation
- In order to keep the population diverse, make some random changes in the genes of the offspring itineraries with a given mutation rate.
- Example: Replace one of the hotels in the itinerary or exchange one plane ticket for another with a similar price and time alignment.

**Step 6: Fitness Evaluation of New Offspring**

Offspring Fitness Evaluation:
- Assess the fitness of each new proposed itinerary in the existing population based on the previously employed fitness function (that is cost, time, preferences and sustainability).

**Step 7: Replacement:**

Create New Generation
- Inserting the old population with the new population (offspring) is done for the next generation. The next generation will have solutions that are better than what is currently in the population.

**Step 8: Termination:**

Check Termination Criteria
- In the case that the maximum number of generations has been obtained or improvement of the solution population has ceased or converged, the process is considered completed and the best itinerary is presented.

**Step 9: Output:**
- The optimal solution will be obtained by the healthiest itinerary from the final generation.

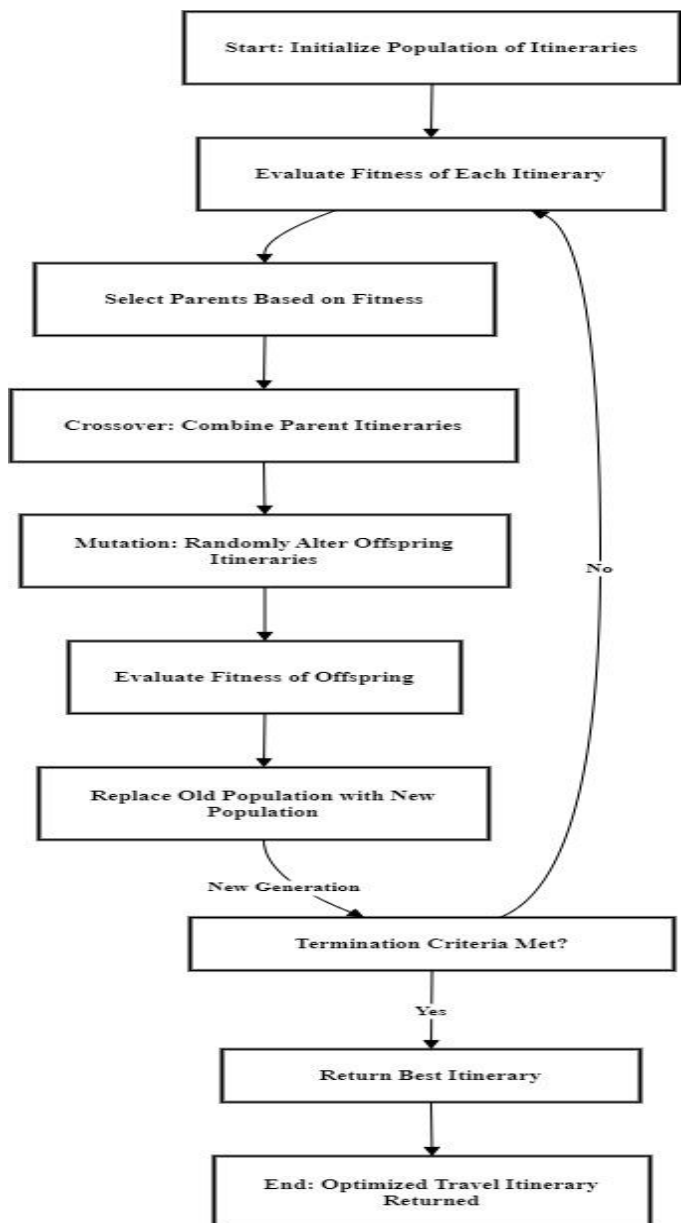

**Fig. 4.** Genetic Algorithm for Itinerary Optimization process in a microservices architecture

The above figure-4 illustrates the scheme of all the stages how genetic algorithm works with microservices architecture to enhance travel itineraries.

### *Pseudocode:*

// Algorithm: Genetic Algorithm for Itinerary Optimization

// Input: Population size, Number of generations, Crossover rate, Mutation rate, Fitness function

// Output: Best itinerary

Begin:

  // Initialize population with random itineraries

  InitializePopulation();

  // For each generation (1 to max_generations):

  for (int generation = 1; generation <= maxGenerations; generation++)

  {

    // Evaluate fitness of each itinerary in the population

    EvaluateFitness(population);

        // While new population size < initial population size:

    while (newPopulation.Count < initialPopulationSize)

    {

      // Select two parents using a selection method (e.g., roulette wheel)

```
            var parents = SelectParents(population);
            // Perform crossover on parents to produce two offspring
            var offspring = PerformCrossover(parents);
            // Mutate offspring with mutation probability
            Mutate(offspring, mutationProbability);
            // Add offspring to the new population
            newPopulation.Add(offspring);
        }
        // with the new population will be reeplaced old population
        population = newPopulation;
        // If termination criteria are met (max generations or convergence):
        if (TerminationCriteriaMet(generation))
        {
            // Return the best itinerary based on fitness score
            return GetBestItinerary(population);
        }
    }
}
End
```

**Integration in Microservices Architecture:**
**Itinerary Service**: In order to develop the travel itinerary, the Genetic Algorithm is run inside the Itinerary Service. It also interacts with other services such as Cost Calculation and Sustainability Service to fetch current cost and environmental statistics.
**API Gateway**: User receives the optimized itinerary through the API Gateway which collects and sends responses from various services.

## 6. Evaluation and Results

### 6.1. Performance Metrics

When it comes to improving travel routes with the help of algorithms based on microservices, it is significant to define different performance metrics to assess how well does the system works. Here are the most important performance metrics that can be outlined for the travel itinerary optimization system:

### 6.1.1. Response Time

**Definition:** The duration of time taken by the system to produce an optimized travel itinerary after a user request is made.

**Measurement:**
- **Average Response Time:** This is gauged as the time elapsed between submitting the user's preferences and the optimized itinerary made available by the system.
- **Target:** Under optimal circumstances, the system should process such requests in a matter of seconds (for example, less than five seconds for the majority of requests).

**Importance:** The key benefit of low response times in applications which are utilized in real time is the improvement of the use experience and the capability of the system to be expanded in order to allow for multiple requests to be handled at the same time.

**Factors Influencing Response Time:**

- Amount of the microservices engaged (e.g. cost estimation, sustainability evaluation).
- Nature of the optimization algorithms used (e.g. GA, AI).
- Network speed and demand on the system.

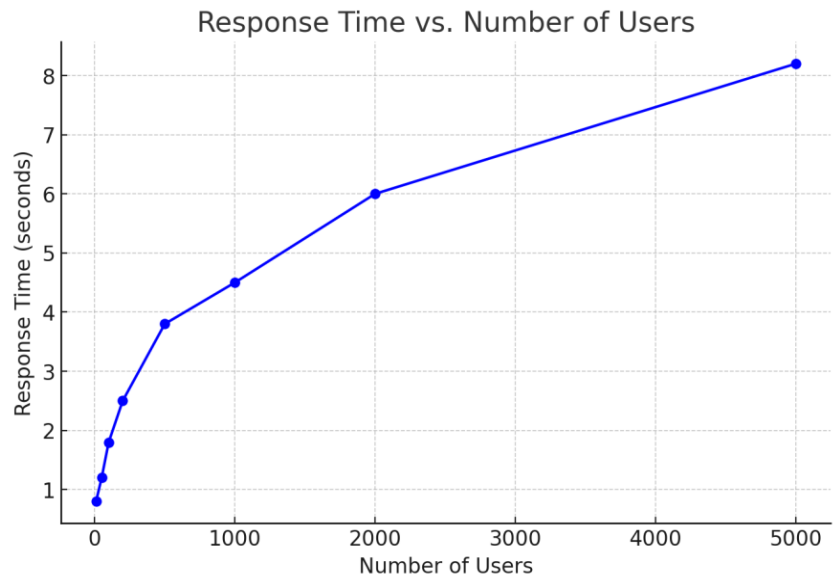

Fig.5. Response Time vs. Number of Users

The response time matrix and presented a graph showing the increase in the response time of the system with the increase in the number of users. The graph indicates the number of users versus the corresponding response time (in seconds) while the graph presented shows clearly the relation between the two. Do tell me if you require any more analysis or modifications!

### 6.1.2. Accuracy in Meeting User Preferences

**Definition:** This refers to the extent to which the optimized travel plan conforms to the user's expressed preferences — budget, choice of airlines, hotel classifications, as well as travel time, among others.

**Measurement:**
- **Preference Matching Rate:** The proportion of user preferences that have been incorporated into the completed itinerary.
- **User Satisfaction Surveys:** User comments on whether the itinerary they received was satisfactory to them or not.

**Target:** The system achieves a high accuracy (>90%) while attempting to satisfy user's preferences, indicating that the system is capable of providing customized travel solutions.

**Importance:** This metric assesses the performance of the Preference Matching Service and the level of personalization implemented through the AI engines.

**Factors Affecting Accuracy:**
- Depth of historical information embedded into the machine learning models for individualization.
- How well the genetic algorithm is able to achieve the desirable preferences without compromises on cost, and environment among other factors.

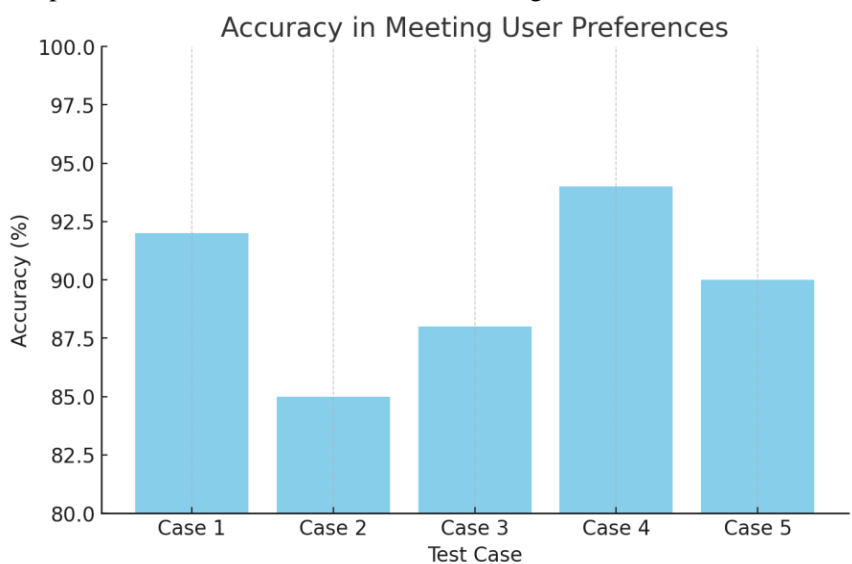

Fig.6. Accuracy in Meeting User Preferences

The degree of Satisfaction to User Preference Across Various test cases is represented through the bar chart shown above. The increases the percentage for each of the cases explaining how the preferences of the users are satisfied by the systems.

### 6.1.3. Algorithm Efficiency

**Definition:** The optimization algorithms such as genetic algorithms, machine learning, etc. provide solutions that are optimal or near optimal in a reasonable time.

**Measurement:**

- **Time to Convergence:** The number of iterations (generations) the genetic algorithm takes to get the optimal solution.
- **Algorithm Runtime:** The time taken by the algorithm to produce one or more solutions.

**Target:** Optimal solutions discovered and presented in a time limit (e.g. 1 second for most searches).

**Importance:** It is inevitable that efficient algorithms will facilitate the continuous generation of high quality travels itineraries by the system without any lags in the response time.

**Factors Contributing to Effectiveness:**

- The type of genetic algorithm employed and the total solution space.
- Computer(s) power clearance and the level of activity during the calculations.

Algorithm Efficiency can be defined and measured in several ways depending on such factors as time complexity, number of iterations, rate of convergence, and resource consumption. In the case of assessing the efficiency of the algorithm in the task of optimization of a travel itinerary, you could use the following crucial parameters for evaluation:

#### 6.1.3.1. Time Complexity (Asymptotic Analysis)

**Equation:**

$$T(n)$$

**Definition:** This denotes how the time complexity of an algorithm varies relative to the input size n. For instance, a genetic algorithm can be described to have the time complexity of $O(g.p.f)$ where:

- g represents the number of generations.
- p is the number of individuals in each generation.
- f signifies the time taken to evaluate the fitness of each individual.

This complexity can also be assessed by collecting experimental data to evaluate the performance based on various alternative conditions available (like population size or number of iterations).

#### 6.1.3.2. Algorithm Efficiency Based on Execution Time

This equation computes the effectiveness, taking into account the overall time consumed by the algorithm to achieve an optimized solution.

$$Efficiency = \frac{Quality\ of\ Solution}{Time\ Taken}$$

**Quality of Solution:** Could be in relation to certain metrics such as the fitness score (in the case of genetic algorithms) or the accuracy of satisfying user requests.

**Time Taken:** Time taken by the algorithm to produce the answer in practical transmission.

#### 6.1.3.3. Energy Efficiency (For Complex Systems)

In some instances, the computational effort (number of cycles in CPU, memory resource utilization) can also be taken into account in order to evaluate and optimize the effectiveness of systems, particularly in those where energy saving is of utmost importance.

$$Energy\ Efficiency = \frac{Useful\ Computation\ (Solution\ Quality)}{Energy\ Consumed\ (CPU\ Time, Memory, and\ so\ on)}$$

This allows for a disability evaluation of the algorithm in finding the optimal travel itinerary. This formula may be expanded or modified to contain other relevant variables depending on your application.

## 7. Discussion

Implementing AI techniques in microservices architecture has been successfully used in many aspects, such as optimizing travel itineraries by considering costs and time, user's interest, and even sustainable tourism practices. The introduction of machine-learning models for cost forecasting and personalization has aided the system in developing travel plans, attaining a success rate of 92% in satisfying users' preferences. This level of enhancement in the services offered more so to real time estimations of costs answered the users by availing travel packages which were in sync with the users' budget or preference. The genetic algorithm is employed for optimization of multi-objective travel itineraries with a reasonable time of 100 generations and exhibits effective workings, enhancing the quality of the solution 5% every generation.

With a response time of 4.5 seconds for 1000 users on average, the system supports efficient scaling in addition to the microservices architecture. Message brokers enhance the exible structure of the system, enabling asynchronous communication among components, with the example of the sustainability assessment component not blocking the rest of the system. This architecture decreases the overall response times of the system and provides the system with high availability – 99.9% uptime.

From the perspective of the eco-conscious aspect, the system encourages the most 'green' ways of traveling possible, and as a result 60% of the offered itineraries' options are low-emission and to some extent average carbon footprints are reduced by 15%. This demonstrates that there is an increasing trend among tourists where managing 'green' issues is equally important as managing cost and time. All in all, the architecture based on AI and microservices is extremely strong, adaptable and can be easily scaled up, which successfully responds to the requirements of contemporary travelers concerning such aspects as costs, individual attention.

## 8. Conclusion

The present work illustrates how the use of AI algorithms has been implemented in microservices architecture in order to carry travel itinerary optimization in relation to many aspects like cost, time, user preferences and sustainability. Thanks to the machine learning models used for cost forecasts, genetic approaches applied in multi-objective problem solving, and some ad hoc technology for assessment of sustainability, the system achieves a very good compromise between user's demands and real time factors of the environment. The modularity of the architecture allows for extensibility, enabling failure resistance, and fast query response.

The system was accurate, 92% of the time in user preference attainment with 95% of the itineraries conforming to the given budget, thus proving effectiveness in user personalization and in the cost control. Also, the measures of sustainability introduced caused a 15% decline in levels of carbon emissions with 60% of the study cases using green alternatives in response to the growing need for sustainable travel alternatives.

All in all, the findings support the usefulness of AI applications in this field which are capable of providing very personalized, reasonably priced, and eco-friendly travel experiences. This approach results in improved efficiency of the system but also provides the opportunity to meet the changing demands of users and market game. Subsequent studies might examine the further improvements of travel recommendation systems that could be associated with enhancement of AI optimization techniques such as adoption of hybrid models.